Magneto-optics of dichroic chiral photonic crystals: resonant diffraction transmission and higher-order reflections at grazing angles

A.H. Gevorgyan[1*], A.A. Malinchenko[1], N.A. Vanyushkin[2], S.S. Golik[1,3], A.V. Davydenko[1], T.M. Sarukhanyan[4], K.B. Oganesyan[5,6]

[1]Far Eastern Federal University, Institute of High Technologies and Advanced Materials, 10 Ajax Bay, Russky Island, Vladivostok 690922, Russia

[2]Moscow Institute of Physics and Technology, Phystech School of Electronics, Photonics and Molecular Physics, Institutsky lane 9, Dolgoprudny, 141700, Russia

[3]Institute of Automation and Control Processes, Far East Branch, Russian Academy of Sciences, 690041 Vladivostok, Russia

[4]Yerevan State University, 1 Alex Manoogian Str., 0025 Yerevan, Armenia

[5]A. Alikhanyan National Lab, Yerevan Physics Institute, Yerevan, Armenia

[6]Institute of Experimental Physics SAS, Kosice, Slovakia

[*]e-mail: agevorgyan@ysu.am

**Abstract**

Magneto-optical effects in structured anisotropic media have attracted significant attention due to their potential in tunable photonic devices. In particular, dichroic cholesteric liquid crystals (i.e., cholesterics in which the real parts of the local dielectric permeability tensor are equal each other, and the imaginary parts differ) provide a unique system in which absorption anisotropy rather than anisotropy of real pars of the local dielectric permeability controls the optical response. In this work, we investigated the magneto-optical properties of dichroic cholesteric liquid crystals under oblique light incidence. To analyze the optical characteristics, we used the Berreman method, which is a matrix approach to solving Maxwell's equations. The unique properties of such structures were revealed, namely, resonant Bragg diffraction transmission instead of diffraction reflection and higher orders of diffraction reflection under oblique light incidence at grazing angles. It was shown that at light oblique incidence in the presence of an external magnetic field, new Dirac points appear in the spectra of such structures at each order of diffraction reflection. Furthermore, it is shown that an external magnetic field suppresses the shift of the polarization of eigenmodes from circular to linear polarization as the angle of incidence increases. This system can be used as a narrowband filter with a widely tunable wavelength.

## 1. INTRODUCTION.

Liquid crystals, first discovered by Reinitzer [1] in 1888 and later named by Lehmann [2] in 1904, represent an aggregate state of matter intermediate between solid crystals and isotropic liquids. They are characterized by anisotropy of optical, electrical, mechanical, and other physical properties, combined with the ability to flow freely, which is inherent in liquids [3–16]. For a liquid crystal, or mesomorphic, state of matter to occur, its molecules must have significant geometric anisotropy. Depending on the structural features of the molecules, when the temperature changes, the substance can sequentially pass through one or more mesophases before reaching the isotropic liquid phase. Of particular interest from the point of view of optical characteristics are mesophases formed by chiral, i.e., mirror-asymmetric molecules. The chirality of a molecule implies the absence of centers of symmetry, inversion axes, and transverse planes in its structure. Materials consisting of such molecules are capable of forming mesophases with a spatially periodic organization, characterized by either right- or left-handed orientation, the so-called chiral liquid crystals. Among the known types of liquid crystals, this category includes cholesteric liquid-crystalline compounds, chiral smectic phases, and the blue phase of cholesterics.

The spatial periodicity of chiral liquid crystal structures, with a period comparable to the wavelength of the optical range, determines their unique optical properties associated with the phenomenon of diffraction. Among the characteristic effects are selective light reflection, giant polarization rotation, and many other phenomena. Under the influence of external factors, such as electric and magnetic fields, heat, light, etc. the optical characteristics of these materials can change significantly [3,9,10,14,17-21]. This makes liquid crystals, especially chiral ones, promising for a wide range of applications. Due to the diversity of observable physical phenomena and their high practical significance, liquid crystals have become the subject of intensive research. They are used in display technologies, microwave radiation detection devices, medical diagnostics, and many other fields. As a result, liquid crystal optics has emerged as an independent field of science, as evidenced by the extensive scientific literature, including a large number of articles, reviews, and monographs [3–21] (see also the extensive literature cited therein).

To date, the optical properties of cholesteric liquid crystals, or simply cholesterics, have been studied in great detail, both theoretically and experimentally. Cholesteric liquid crystals (CLCs) represent a liquid crystal phase and can consist either of a single material composed of chiral anisotropic molecules or of mixtures of one or more chiral additives in a nematic liquid crystal (NLC), resulting in the formation of a spatial helical structure [3,9,10,14,17-21]. In each layer, the orientation of the long axes of the molecules remains the same and perpendicular to the common optical axis (helix axis) of the system. At the same time, there is no long-range order in the arrangement of the centers of mass of the molecules in each layer. The direction of the molecular axes in subsequent layers gradually rotates relative to the previous one by a certain angle, ensuring a periodic spiral organization.

This periodicity leads to the emergence of a photonic band gap (PBG) that is sensitive to light polarization. The spatial period of the helix, or the pitch of the helical structure, lies in the optical range, and its value, like the direction of the local optical axis, can change under the influence of external factors such as pressure, temperature, electromagnetic fields, the presence of impurities, and surface effects [22–42]. The unique properties of CLCs allow the creation of a wide variety of optical elements and devices based on them. In particular, media for photo-optical data recording [43–46], sensors [45,47-49], electrically switchable mirrors [50,51], optically controlled linear polarization rotators [52], coatings with controlled friction and adhesion [53], materials for display technology [44,54], materials for low-threshold lasing [55-58], reliable authentication [59], metal ion sensors [60], materials for switchable smart windows [61], and many others [62].

In recent years, magneto-optics of cholesterics, which studies changes in the optical properties of CLCs under the influence of an external magnetic field, has attracted considerable interest. The influence of an external magnetic field on CLCs can lead to a variety of effects; in particular, it can change the helix pitch, change the texture and conductivity, lead to surface effects in thin layers, transform irregular and heterogeneous samples into a regular helical structure, and lead to the transition of a planar structure into a confocal structure. Problems concerning the study of the Faraday effect in CLCs have become very relevant in recent years and have been addressed in many works [63-83] (see also the extensive literature cited therein).

This work examines the optical properties of dichroic cholesteric liquid crystals, i.e., cholesterics in which the real parts of the principal values of the local dielectric permeability tensor coincide, while the imaginary parts differ (dichroic cholesterics).

The present study focuses on the specific features of the influence of a magnetic field on dichroic cholesterics. Although the formation of the helical structure of CLCs is usually associated with the presence of local dielectric anisotropy, in this case, a model in which this anisotropy is absent is considered. This allows us to exclude background effects and clearly identify the changes caused by the magnetic field at oblique incidence, as described earlier in [84].

In addition, the magnetic field may have a deforming effect on the structure of the CLC itself, but since this analysis considers the absence of local dielectric or magnetic anisotropy, and the magnetic field is directed along the axis of the helix, this effect can be neglected.

## 2. METHODS

Let us assume that the tensors of dielectric and magnetic permeability of magnetoactive dichroic CLC (Fig. 1) have the form:

$$\hat{\varepsilon}(z) = \varepsilon_m \begin{pmatrix} 1 + \delta\cos 2az & \pm\delta\sin 2az \pm \frac{ig}{\varepsilon_m} & 0 \\ \pm\delta\sin 2az \mp \frac{ig}{\varepsilon_m} & 1 - \delta\cos 2az & 0 \\ 0 & 0 & 1 - \delta \end{pmatrix}$$

$$\hat{\mu}(z) = \hat{I}, \qquad (1)$$

where $\varepsilon_m = (\varepsilon_1 + \varepsilon_2)/2$, $\delta = (\varepsilon_1 - \varepsilon_2)/(\varepsilon_1 + \varepsilon_2)$, $\varepsilon_{1,2}$ are the principal values of the local dielectric permittivity tensor, here we take $\varepsilon_{1,2} = \varepsilon_0 + i\varepsilon''_{1,2}$ with $\varepsilon_0$ to be the real part of the dielectric tensor components, which is assumed to be the same for all components, g is the parameter of magnetooptical activity, $a = 2\pi/p$, a $p$ is helix pitch, $\hat{I}$ is the unit matrix.

Note that, to our knowledge, homogeneous dielectric media with parameters $\text{Re}\varepsilon_1 = \text{Re}\varepsilon_2 = \text{Re}\varepsilon_3$ and absorption anisotropy $\text{Im}\varepsilon_1 \neq \text{Im}\varepsilon_2 \neq \text{Im}\varepsilon_3$ do not exist, however, media with isotropy of refractive index and anisotropic absorption can be created artificially based on metamaterials (note that media with parameters $\text{Re}\varepsilon_1 = \text{Re}\varepsilon_2 = \text{Re}\varepsilon_3$ and $\text{Im}\varepsilon_1 \neq \text{Im}\varepsilon_2 \neq \text{Im}\varepsilon_3$ are also media with effective refractive anisotropy). Indeed, as our calculations show, for example, helical media with double (both dielectric and magnetic) anisotropy and with parameters $\text{Re}\varepsilon_1 \text{Re}\mu_1 = \text{Re}\varepsilon_2 \text{Re}\mu_2$ and $\text{Im}\varepsilon_1 \neq \text{Im}\varepsilon_2$, $\text{Im}\mu_1 = \text{Im}\mu_2 = 0$ possess the full set of properties of media (1) with $\varepsilon_{1,2} = \varepsilon_0 + i\varepsilon''_{1,2}$. The latest achievements in the field of nanostructuring of chiral soft and hard PCs, including those based on new materials [85-101], suggest that the creation of helical structures with such parameters is quite realistic.

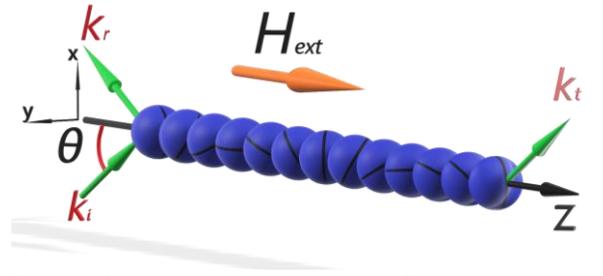

Fig. 1. Geometry of the problem. The blue spheres represent isotropic CLC molecules without birefringence, the black lines in these spheres correspond to the directions of the absorption oscillators, these directions change continuously, forming a helical structure along the z-axis. $H_{\text{ext}}$ is the external static magnetic field. The diagram shows one period of rotation around the crystal helix axis, $k_i$, $k_r$ and $k_t$ are the wave vectors of the incident, reflected, and transmitted waves, respectively.

An external magnetic field can cause not only the Faraday effect, but also directly influence the values of the real parts of the dielectric permeability tensors $\text{Re}\varepsilon_{1,2}$, and this influence has a quadratic dependence on the field strength (see, for example, [102]). In this study, at the first stage, we do not take into account possible changes in $\text{Re}\varepsilon_{1,2}$.

The analysis of light propagation is also carried out under the assumption that we do not consider effects related to frequency dispersion of dielectric or magnetic permeability, as well as absorption dispersion. Thus, all components of the tensors remain constant and independent of frequency; a similar assumption is made for their imaginary parts $\text{Im}\varepsilon_{1,2}$. It should also be noted that the optical properties of such structures were also considered in articles [71,103,104].

The Berreman method, which is a matrix approach to solving Maxwell's equations, was used to analyze the optical characteristics of CLCs. This numerical method is effectively used to simulate light propagation in media with one-dimensional inhomogeneity, where the parameters of optical anisotropy vary only along one spatial direction—in this case, along the helix axis of the CLC structure oriented along the z-axis.

And as in [106], we assume that the magneto-optical activity parameter g is directly proportional to the strength of the external magnetic field: $g = fH_{ext}$ with a proportionality coefficient $f$.

According to the Berreman approach, it is convenient to transform Maxwell's equations into a system of four coupled first-order differential equations written in standard notation as follows:

$$\frac{\partial}{\partial z}\vec{X} = ik_0 \hat{A}\vec{X}, \quad (2)$$

where $\hat{A}$ is a matrix of dimension 4×4 which depends on $\hat{\varepsilon}(z)$, and $\hat{\mu}(z)$ and wave numbers $k_x$, and $k_y$, the elements of which were obtained in [54], $\vec{X} = \{E_x, E_y, H_x, H_y\}^T$ is a four-component column vector. In general, equation (2) can only be solved numerically, and its solution can be represented as follows:

$$\vec{X}(z) = \hat{M}(z)\vec{X}(0) \quad (3)$$

where the matrix $\hat{M}$ is called a transfer matrix, which connects the fields at two points in space. The main convenience of transfer matrices is that the matrix of a complex system consisting of many different elements is easily defined as the product of the matrices of all these elements.

Next, it is convenient to move from the x- and y- components of the E and H fields to the full amplitudes of the electric field E of four waves of different polarization (s- and p-) propagating in opposite directions (+ and −). Let us denote the transition matrix between the two representations by $\hat{S}$:

$$\vec{X} = \begin{pmatrix} E_x \\ E_y \\ H_x \\ H_y \end{pmatrix} = \begin{pmatrix} \cos\theta & \cos\theta & 0 & 0 \\ 0 & 0 & 1 & 1 \\ 0 & 0 & -n_s\cos\theta & n_s\cos\theta \\ n_s & -n_s & 0 & 0 \end{pmatrix} \begin{pmatrix} E_+^p \\ E_-^p \\ E_+^s \\ E_-^s \end{pmatrix} = \hat{S}\vec{\Psi}, \quad (4)$$

where $n_s$ is the refractive index of the medium surrounding the CLC layer, $\hat{S}$ is the transition matrix from the x- and y- components of the E and H fields to the total amplitudes of the electric field strength E of four waves of s- and p- polarization.

To solve the scattering problem, we use the matrix equation for the fields at the two boundaries of the layer [105]

$$\vec{\Psi}_t = \hat{S}^{-1}\hat{M}\hat{S}(\vec{\Psi}_0 + \vec{\Psi}_r) = \hat{m}(\vec{\Psi}_0 + \vec{\Psi}_r), \quad (5)$$

Here $\vec{\Psi}_0 = (E_0^p, 0, E_0^s, 0)^T$, $\vec{\Psi}_r = (0, r_p E_0^p, 0, r_s E_0^s)^T$ and $\vec{\Psi}_t = (t_p E_0^p, 0, t_s E_0^s, 0)^T$ denote the field vectors of the incident, reflected and transmitted waves, and $r_{s,p}$ and $t_{s,p}$ are the reflection and transmission coefficients of the waves of the corresponding polarizations.

Bloch waves were also investigated in this work and calculated from the characteristic equation:

$$\det[\hat{M} - e^{iKp}\hat{I}] = 0, \quad (6)$$

where K is the Bloch wave vector.

To study the features of the light field intensity distribution, we modify formula (5) by solving the matrix equation along the z axis of the crystal:

$$\vec{\Psi}(z) = \hat{S}^{-1}\hat{M}(z)\hat{S}(\vec{\Psi}_0 + \vec{\Psi}_r) = \hat{m}(z)(\vec{\Psi}_0 + \vec{\Psi}_r), \quad (7)$$

The resulting vector $\vec{\Psi}(z)$ contains all the necessary components of the fields for calculating the intensity $I$ according to the formula (8):

$$I = I_p + I_s = |E_+^p + E_-^p|^2 + |E_+^s + E_-^s|^2, \quad (8)$$

Using the numerical method of the Berreman transfer matrix, we can solve the problem of reflection, transmission, and absorption of light and etc. in the case of a planar CLC layer of finite thickness under both normal and oblique incidence of light. We assume that the optical axis of the CLC layer is perpendicular to the boundaries of the layer and directed along the z-axis. The CLC layer is bordered on both sides by isotropic half-spaces with the same refractive indices equal to $n_s$.

## 3. RESULTS AND DISCUSSION

First, we consider the case of no external magnetic field. Next, we consider the structure of a dichroic CLC with a right-handed helix, layer thickness $d=50p$, and helix pitch $p = 420$ nm. The real part of the local dielectric permeability tensor CLC is equal to $\varepsilon_0 = 2.25$. Here and further, two cases are considered: the case of minimal influence of dielectric boundaries, i.e., $n_s = \sqrt{\varepsilon_0}$, and the case when the layer is in a vacuum, i.e., when $n_s = 1$.

Next, each optical system has characteristic polarizations called eigenpolarizations (EP). EPs are two polarizations of incident light that do not change when light passes through the system. They coincide with the polarizations of the eigenmodes excited in the medium. The EPs of the CLC layer at normal incidence of light practically coincide with orthogonal circular polarizations, although in general they can differ significantly from circular ones. Let us enumerate the EPs as follows: we will assume that the first EP is an EP that approximately coincides with the right circular polarization and diffracts on the periodic structure of the CLC (under normal incidence of light), and the second EP coincides with the left circular polarization and does not diffract on the periodic structure of the CLC (again, at normal incidence). At oblique incidence, both polarizations become diffracting, but incident light with polarization coinciding with the first EP is strongly diffracting, and with the second, weakly diffracting. EPs shift from circular to linear polarization as the angle of incidence increases.

### 3.1. Resonant Bragg diffraction transmission and higher orders of diffraction reflection at oblique incidence at grazing angles.

Fig. 2 shows the transmission spectra $T = |E_t|^2/|E_i|^2$ (first column), reflection spectra $R = |E_r|^2/|E_i|^2$ (second column), and absorption spectra $A = 1 - (R + T)$ (third column) at angles of incidence $\theta = 0°$ (first row), $\theta = 30°$ (second row), $\theta = 60°$ (third row), $\theta = 80°$ (fourth row), $\theta = 85°$ (fifth row) and $\theta = 89°$ (sixth row) in the case when the medium layer is in a vacuum, i.e., when $n_s = 1$. Here, $\vec{E}_i$, $\vec{E}_r$, $\vec{E}_t$ are the incident, reflected, and transmitted fields, respectively. The solid lines correspond to the case $\text{Im}\varepsilon_1 = 0.2$ and $\text{Im}\varepsilon_2 = 0$, while the dashed lines correspond to the case $\text{Im}\varepsilon_1 = 0.$ and $\text{Im}\varepsilon_2 = 0.2$. In Fig. 2, the red lines correspond to incident light with polarization coinciding with the first EP, and the blue lines correspond to the second EP.

First, let us consider the characteristics of the reflection, transmission, and absorption spectra in the case of $\text{Im}\varepsilon_1 = 0.2$ and $\text{Im}\varepsilon_2 = 0$. With normal incidence of light, we have a resonance peak of diffraction reflection at the Bragg wavelength $\lambda_B = p\sqrt{\varepsilon_0}$, i.e., in this case, there is no extended region of diffraction reflection with a finite frequency width, and there is a resonance peak of reflection at the Bragg wavelength. Another feature of dichroic CLCs is that there is also a resonance peak (albeit significantly lower) at this wavelength in the transmission spectrum. As the angle of incidence increases, the Bragg peak experiences a blue shift. At small angles of incidence, the changes in the height of these peaks are relatively insignificant. However, starting at $\theta = 60°$, as the angle of incidence continues to increase, the height of the transmission peak increases and the reflection peak decreases. At $\theta = 89°$, there is practically no reflection (at large angles, the reflection peak was replaced by a dip in the reflection spectrum at the Bragg frequency), while at the same time there is a strong resonant increase in the height of the transmission peak at the Bragg frequency. Thus, at large angles of incidence, the Bragg diffraction reflection is replaced by Bragg diffraction transmission.

Thus, we demonstrate the existence of a narrowband transmission line in the 465-475 nm range (bandwidth 0.2 nm) with a transmission of approximately T ~ 0.55. This result demonstrates the possibility of achieving tunable and highly selective filtering in the short-wavelength range of visible light and is very attractive for photonic applications.

Now we need to identify the mechanism of resonant diffraction transmission in an absorbing medium. To do this, we will investigate the peculiarities of the polarization characteristics and the peculiarities of the localization of the total wave excited in the medium.

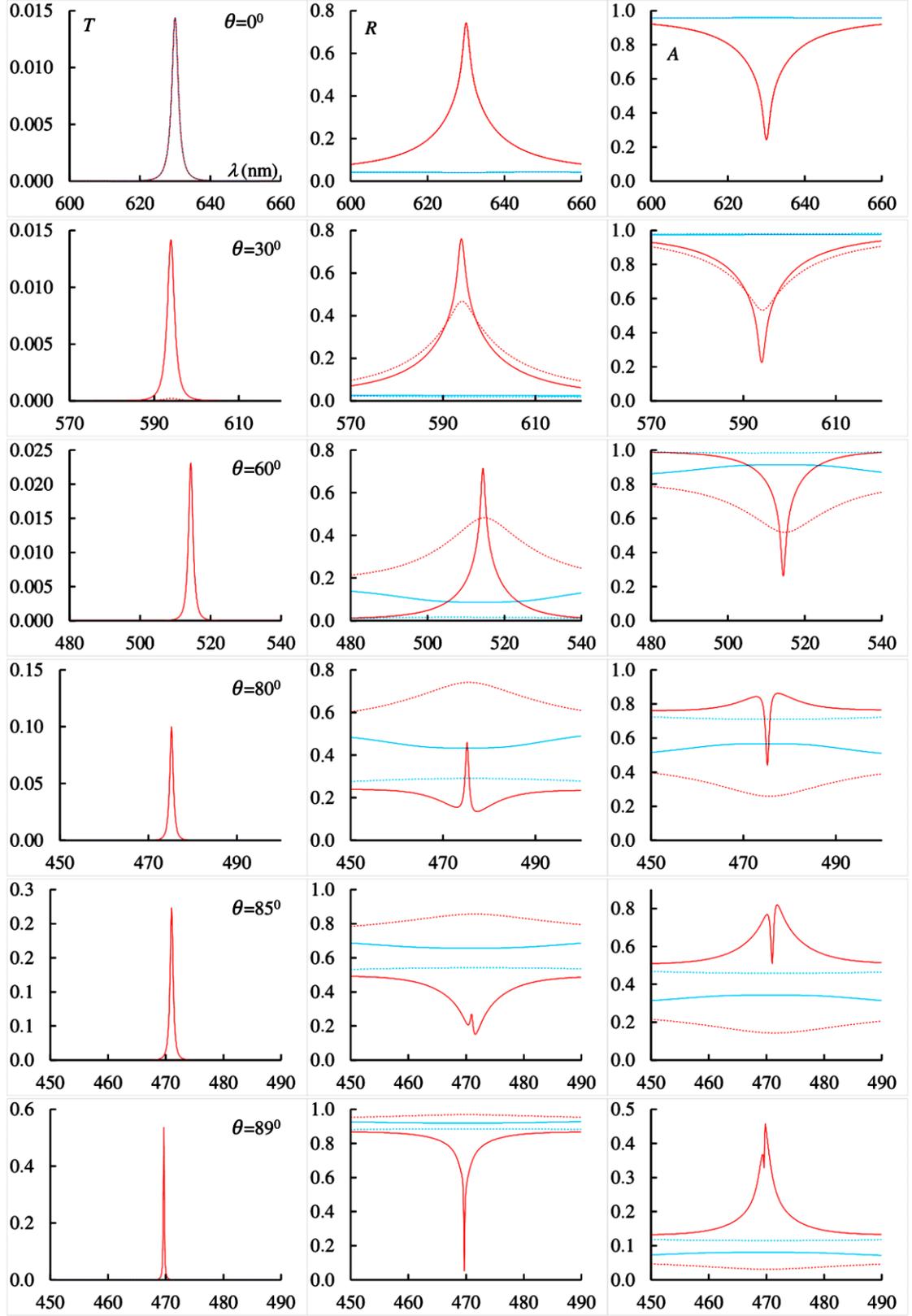

Fig. 2. Transmission spectra $T$ (first column), reflection spectra $R$ (second column), and absorption spectra $A$ (third column) at angles of incidence $\theta = 0°$ (first row), $\theta = 30°$ (second row), $\theta = 60°$ (third row), $\theta = 80°$ (fourth row), $\theta = 85°$ (fifth row), and $\theta = 89°$ (sixth row) in the case when the medium layer is in a vacuum, i.e., at $n_s = 1$. The red lines correspond to incident light with polarization coinciding with the first EP, and the blue lines correspond to the second EP. The solid lines correspond to the case $\mathrm{Im}\varepsilon_1 = 0.2$ and $\mathrm{Im}\varepsilon_2 = 0$, while the dashed lines correspond to the case $\mathrm{Im}\varepsilon_1 = 0.$ and $\mathrm{Im}\varepsilon_2 = 0.2$.

Fig. 3 shows the dependence of the azimuth $\varphi_{in} = \frac{1}{2}\mathrm{arctg}\left(\frac{2\mathrm{Re}\chi_{in}}{1-|\chi_{in}|^2}\right)$ (the angle between the direction of total electric field and $y$-axis; red line), ellipticity of polarization $e_{in} = tg\left(\frac{1}{2}\arcsin\left(\frac{2\text{Im}\chi_{in}}{1+|\chi_{in}|^2}\right)\right)$ (blue line) and the intensity $I(z)$ (green line) of the total wave exited in the medium on the coordinate $z$. Here $\chi_{in} = E_{in}^p/E_{in}^s$, and $E_{in}^{p,s}$ are the $p$ and $s$ components of the electric field of the total wave exited in medium. The incident wave has polarization coinciding with the first EP. Note that although these dependencies are presented in the interval $0 \leq z \leq 2p$ for clarity, these patterns are observed along the entire length of the helix axis $0 \leq z \leq 50p$.

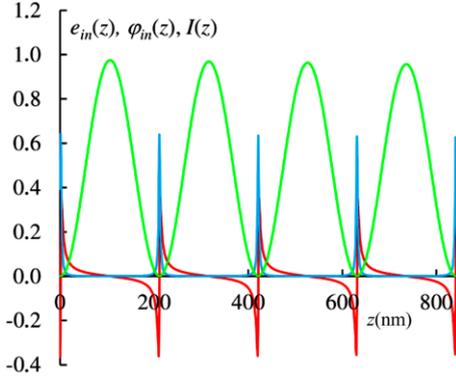

Fig. 3. Dependence of the azimuth $\varphi_{in}$ (red line), ellipticity $e_{in}$ (blue line), and intensity $I$ (green line) of the total wave excited in the medium on the coordinate $z$. The incident wave has a polarization coinciding with the first EP.

As can be seen from the results presented, the total field is coiled to the cholesteric structure, being oriented mainly parallel to the direction of minimum (zero) absorption. The ellipticity of this wave is mainly zero and has narrow peaks near the points $z = i\frac{p}{2}, i = 1,2,\ldots,100.$. This field is mainly localized between the points $i\frac{p}{2}$ and $(i \pm 1)\frac{p}{2}$ and becomes zero at the points $i\frac{p}{2}$. These features determine the possibility of resonant diffraction transmission at large angles of incidence. Note that complete suppression of absorption does not occur because this field does not coil perfectly around the cholesteric structure and the ellipticity is not zero everywhere. For example, in conventional cholesterics with refractive anisotropy under certain conditions (namely, at $n_s = \sqrt{\varepsilon_m}$, at normal incidence, at anisotropic absorption with $\frac{\text{Im}\varepsilon_1+\text{Im}\varepsilon_2}{2} = \pm\frac{\text{Im}\varepsilon_1-\text{Im}\varepsilon_2}{2}$, for incident light with a diffracting EP), the total field in the PBG has linear polarization and, in the presence of anisotropic absorption at one of the PBG boundaries, it coils to a helix with an ideal linear law $\varphi(z) = \frac{2\pi}{p}z$, remaining parallel to the direction of minimum absorption everywhere (here, the absorption suppression effect (Bormann effect) is observed, and near the other boundary it coils to a helix, remaining parallel to the direction of maximum absorption everywhere (here the effect of anomalously strong absorption is observed) [70]. Note that nonlinearity in the dependence $\varphi$ on $z$ appears already at $n_s \neq \sqrt{\varepsilon_m}$.

For comparison, Fig. 4 shows the same dependencies as in Fig. 3 for the mode with weakly diffracting EP at the same angle of incidence $\theta = 89°$.

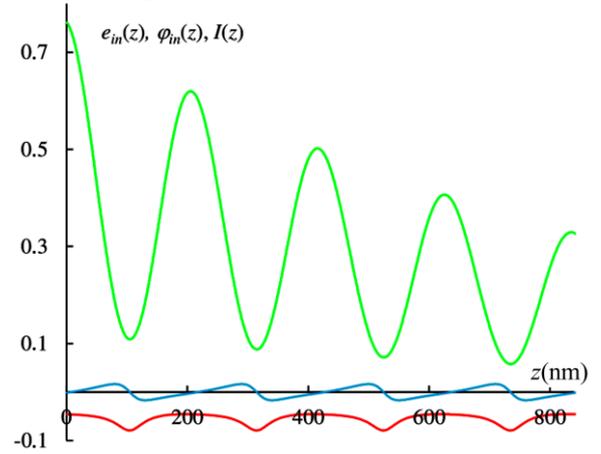

Fig. 4. Same as in Fig. 3 dependencies for the mode with the second EP.

Next, Fig. 5 shows the dependencies of azimuth $\varphi_{in}$ and ellipticity $e_{in}$, and Fig. 6 shows the dependence of $I$ on wavelength at $z = \frac{p}{4}$. As can be seen from these graphs, the azimuth $\varphi_{in}$ and ellipticity $e_{in}$ turn to zero, and $I$ has a resonance peak exactly at the wavelength of the resonance diffraction transmission.

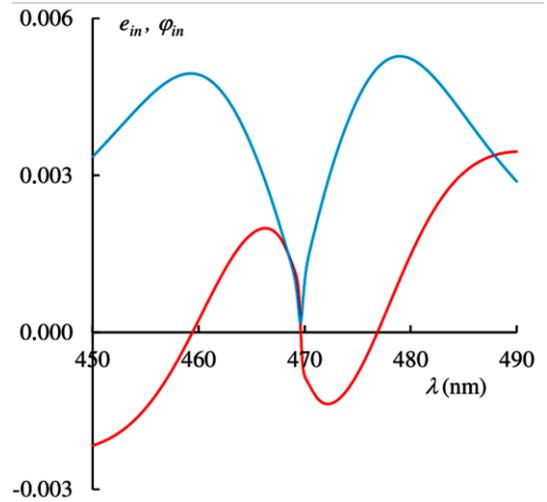

Fig. 5. Dependence of the azimuth $\varphi_{in}$ (red line) and ellipticity $e_{in}$ (blue line) on wavelength at $z = \frac{p}{4}$. The incident wave has polarization coinciding with the first EP.

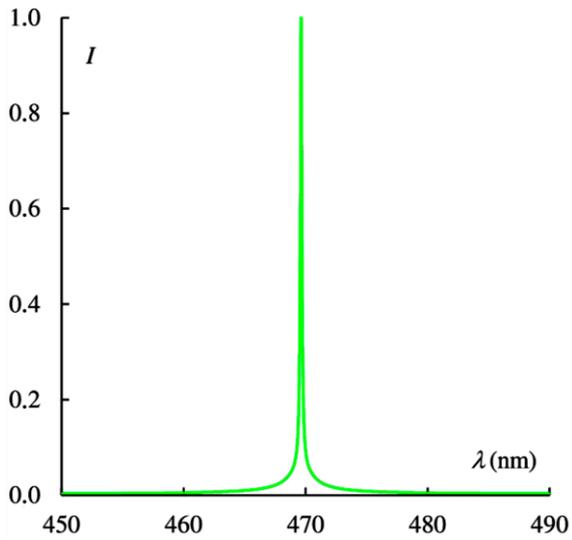

Fig. 6. Dependence of intensity $I$ on wavelength at $z = \frac{p}{4}$. The incident wave has polarization coinciding with the first EP.

For comparison, in Fig. 7 we present the same dependencies as in Figs. 5 and 6 for the mode with the second EP.

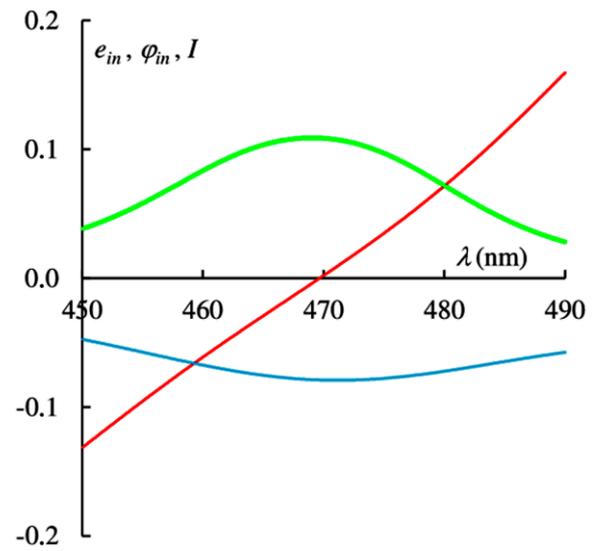

Fig. 7. Same as in Figures 5 and 6, dependencies for the mode with the second EP.

Let us return to Fig. 2. As can be seen from this figure, in the case of $\text{Im}\varepsilon_1 = 0$ and $\text{Im}\varepsilon_2 = 0.2$, the described effect is not observed. This is natural, since in this case, when light falls obliquely in a plane transverse to the direction of light propagation, there is no direction with zero absorption.

Fig. 8 shows the same spectra as in Fig. 2 at $n_s = \sqrt{\varepsilon_0}$, i.e., in the case of minimal influence of dielectric boundaries.

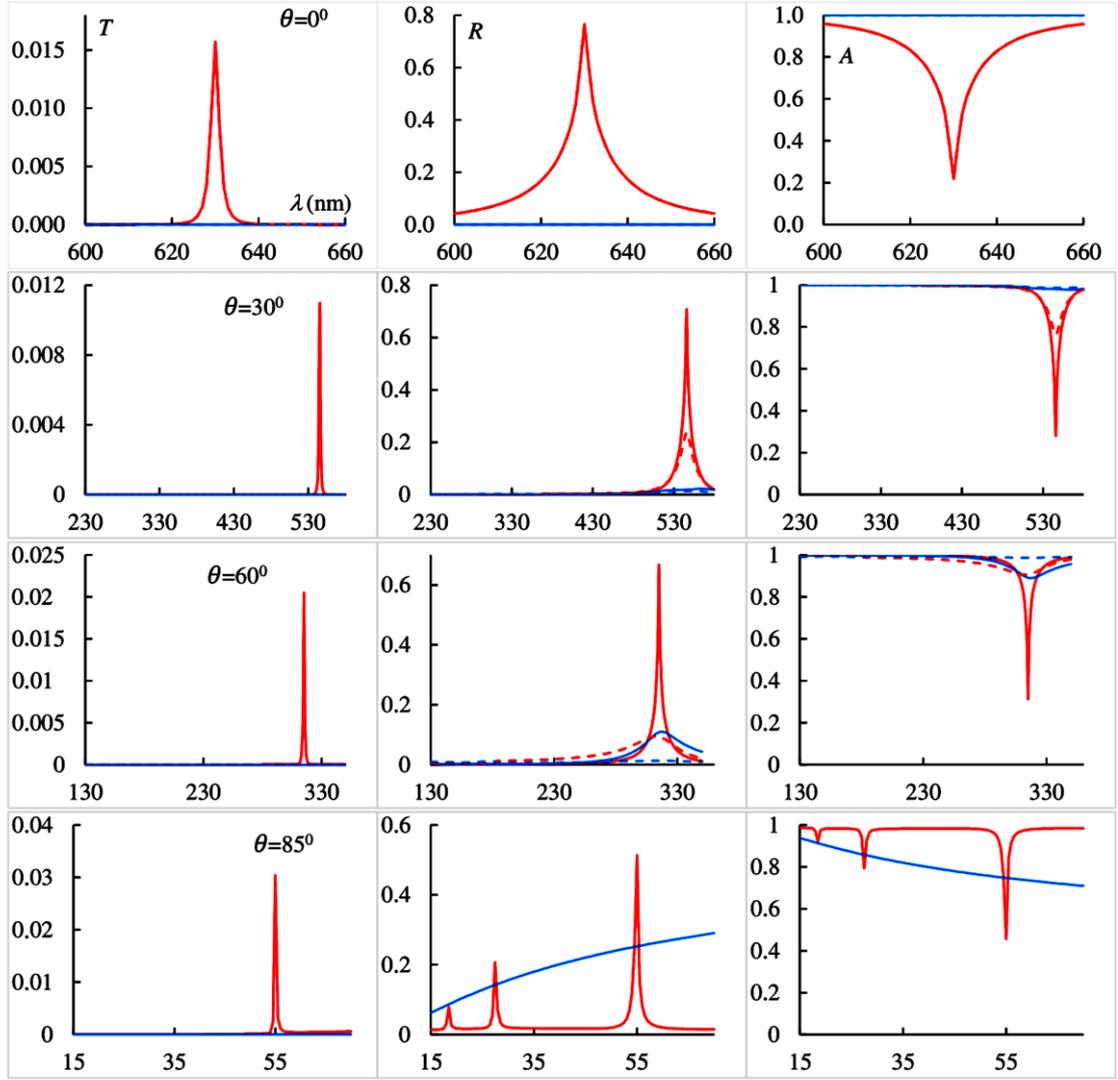

Fig. 8. The same spectra as in Fig. 2 at $n_s = \sqrt{\varepsilon_0}$.

As can be seen from the graphs presented, at large angles of incidence with $\text{Im}\varepsilon_1 = 0.2$ and $\text{Im}\varepsilon_2 = 0$, in this case there is also an increase in transmission and a decrease in reflection at the Bragg frequency, but at $n_s = \sqrt{\varepsilon_0}$ these changes are insignificant. However, in this case, another unique effect is observed. As can be seen from this figure, at an angle of incidence $\theta = 85°$, higher orders of diffraction reflection are excited. At smaller angles of incidence, they are also excited, but much weaker. The uniqueness of this effect lies in the fact that it is observed in a strongly absorbing medium. Note that at $\text{Im}\varepsilon_1 = 0$ and $\text{Im}\varepsilon_2 = 0.2$, higher orders of diffraction reflection are not observed at any angles of incidence; moreover, in this case, at large angles of incidence, large imaginary parts of the wave vectors are obtained, and the reflection and transmission spectra cannot be calculated (see the spectra in Fig. 8 at an angle of incidence $\theta = 85°$).

Now let us examine the evolution of the reflection $R$ and transmission $T$ spectra as the angle of incidence of light changes to get the most complete picture of the characteristics of these spectra. Fig. 9 shows the evolution of the reflection $R$ (first row) and transmission $T$ (second row) spectra as the angle of incidence of light $\theta$ changes at $n_s = 1$. The light incident on the layer has right (left column) and left (right column) circular polarization.

Fig. 10 shows the same evolution of the reflection $R$ and transmission $T$ spectra as in Fig. 9 when the angle of incidence $\theta$ is changed, this time at $n_s = \sqrt{\varepsilon_0}$.

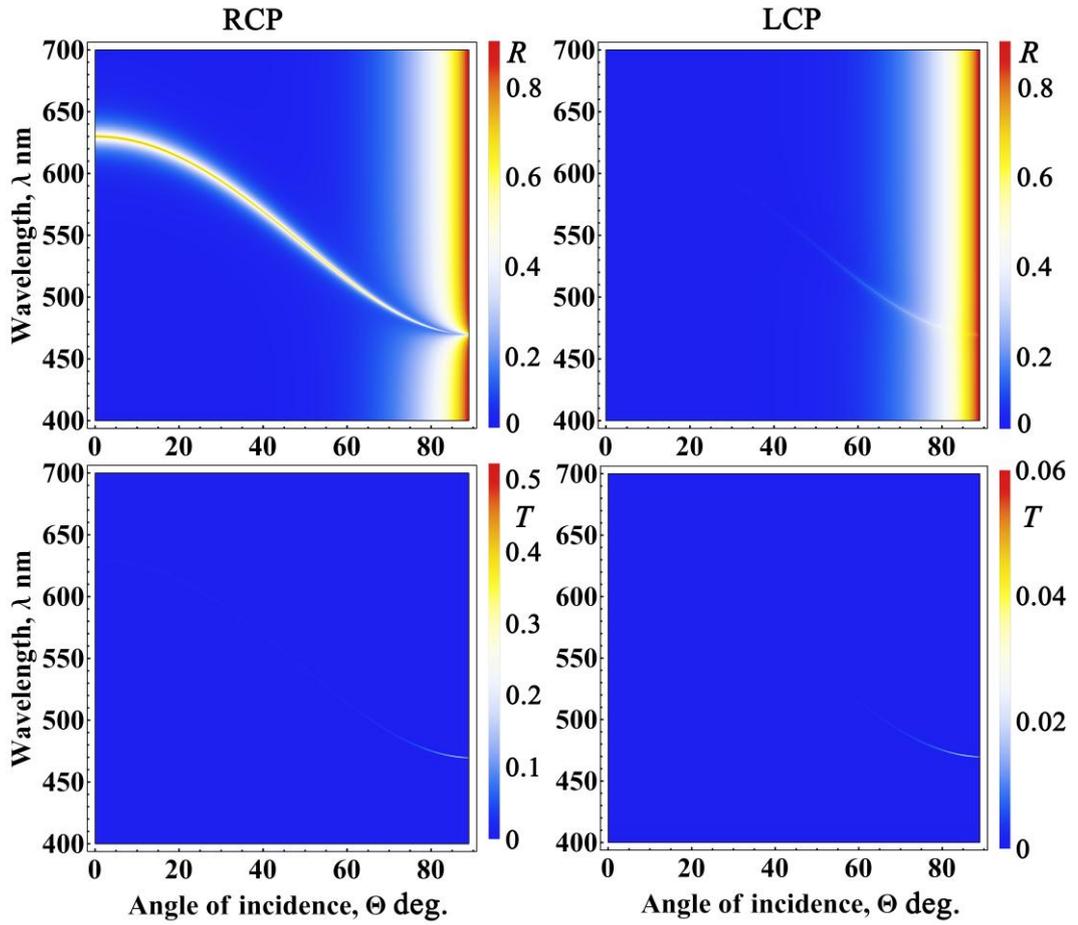

Fig. 9. Evolution of reflection spectra $R$ (first row) and transmission spectra $T$ (second row) as the angle of incidence $\theta$ changes at $n_s = 1$. Light incident on the layer has right circular polarization (left column) and left circular polarization (right column).

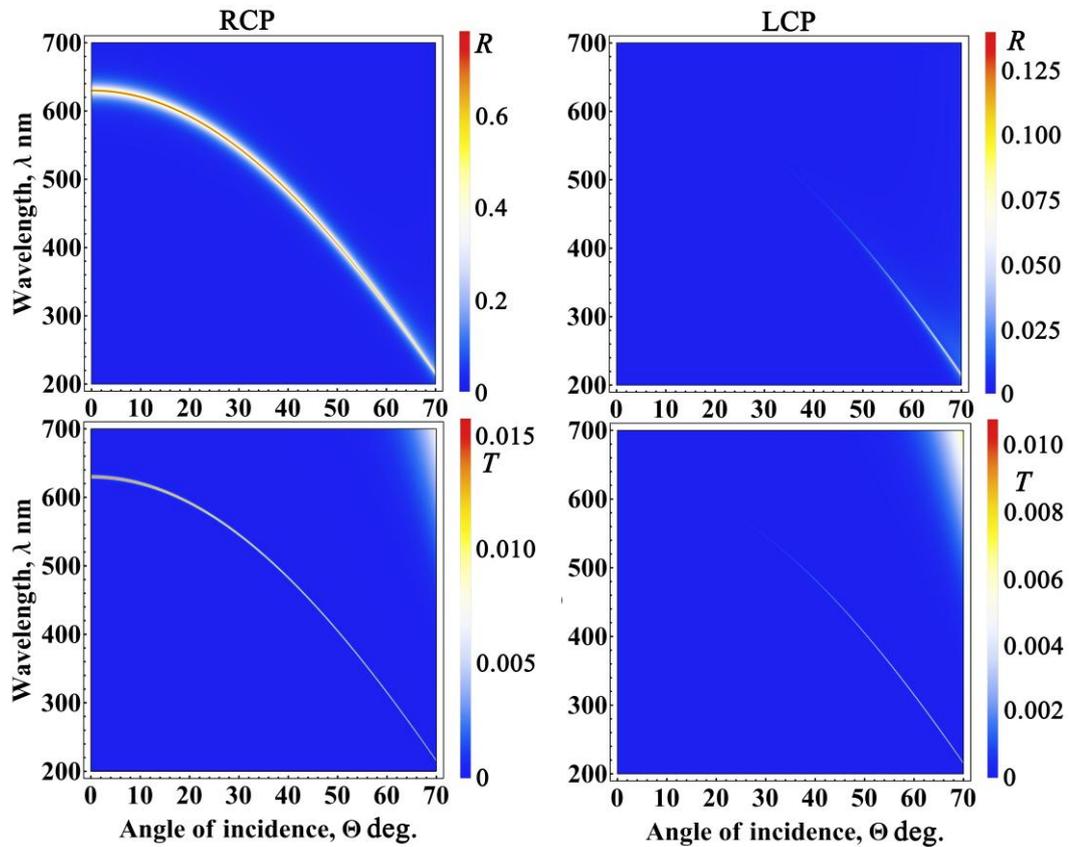

Fig. 10. Same as in Fig. 9, evolution of reflection $R$ and transmission $T$ spectra with changing angle of incidence $\theta$, this time at $n_s = \sqrt{\varepsilon_0}$

In conclusion of this paragraph, we note that when $\text{Re}\varepsilon_1 = \text{Re}\varepsilon_2$ and $\text{Im}\varepsilon_1 \neq \text{Im}\varepsilon_2$, the medium is still locally birefringent, since $\text{Re}n_1 = \text{Re}\sqrt{\varepsilon_1} \neq \text{Re}n_2 = \text{Re}\sqrt{\varepsilon_2}$. However, if we choose $\text{Re}\varepsilon_2$ such that $\text{Re}n_1$ becomes equal to $\text{Re}n_2$, the Bragg diffraction reflection decreases but does not disappear.

**3.2. New magnetically induced Dirac points.** Now, we move on to studying the magneto-optical properties of dichroic cholesterics. First, we will investigate the properties of Bloch wave vectors excited in magnetically active dichroic cholesteric in an external magnetic field at an angle to the axis of light propagation.

Note that the analysis of Bloch wave spectra is very important in studying the dispersion properties of periodic structures. Bloch wave spectra reveal important features of the band structure, such as the presence of PBGs, Dirac points, and exceptional points, and also provide significant insight into the photonic properties of periodic media. Recall that in the absence of absorption and amplification in a photonic crystal, the real Bloch wave vectors $K_z$ correspond to propagating modes, while the imaginary $K_z$ correspond to evanescent modes or PBG. It is known (see, for example, [69]) that, in the presence of an external magnetic field directed along the helix axis, when light propagates along the helix axis of the liquid crystal, the wave vectors of the liquid crystal's eigenwaves shift not only along the wavelength axis (they undergo a blue shift), but also perpendicular to this axis. This leads to the possibility of wave vectors intersecting and the emergence of new Dirac points, as well as the emergence of exceptional points [107].

Fig. 11 shows the dependencies of the real (solid lines) and imaginary (dashed lines) parts of Bloch wave vectors on wavelength in the absence of an external magnetic field, i.e., at g=0 (first row), in the presence of an external magnetic field at g=1 (second row) and at g=−1 (third row) in the case of light propagation at an angle $\vartheta = 45°$ relative to the axis of the medium. The right column corresponds to dichroic CLC. For comparison, the left column shows the same dependencies for conventional CLC with parameters $\text{Re}\varepsilon_1 = 2.29$, $\text{Re}\varepsilon_2 = 2.143$, $\text{Im}\varepsilon_1 = \text{Im}\varepsilon_2 = 0$. The red and green lines correspond to resonant wave vectors, while the blue and purple lines correspond to non-resonant wave vectors.

As can be seen from Fig. 11, the presence of an external magnetic field leads to the emergence of new lateral Dirac points (intersections of the real parts of Bloch wave vectors) due to the above-mentioned displacement of wave vectors in a direction perpendicular to the wavelength axis (see Fig. 11, second and third rows; in the second row, these points are indicated by arrows). The graphs also show that:

1. For ordinary cholesterics at the absence of external magnetic field the real parts of the resonant wave vectors are identically equal to zero in the PBG region, and here their imaginary parts are different from zero, including in the absence of absorption. When g=1, the real parts of the resonant wave vectors (red and green solid lines) shift upward, while the non-resonant ones shift downward.

2. At g = −1, we have the opposite picture, i.e., the real parts of the resonant wave shift downward, while the non-resonant ones shift upward.

3. For conventional CLCs in the absence of absorption, there is no shift in the imaginary parts of the wave vectors, i.e., the imaginary parts caused by the periodic structure of the medium do not shift, while the imaginary parts of the wave vectors of dichroic CLCs also shift.

4. Around the lateral Dirac points in conventional liquid crystals, regions with nonzero imaginary parts for two of the four wave vectors are formed in oblique incidence, whereas for dichroic liquid crystals, the imaginary parts of the wave vectors do not undergo noticeable changes near the Dirac points.

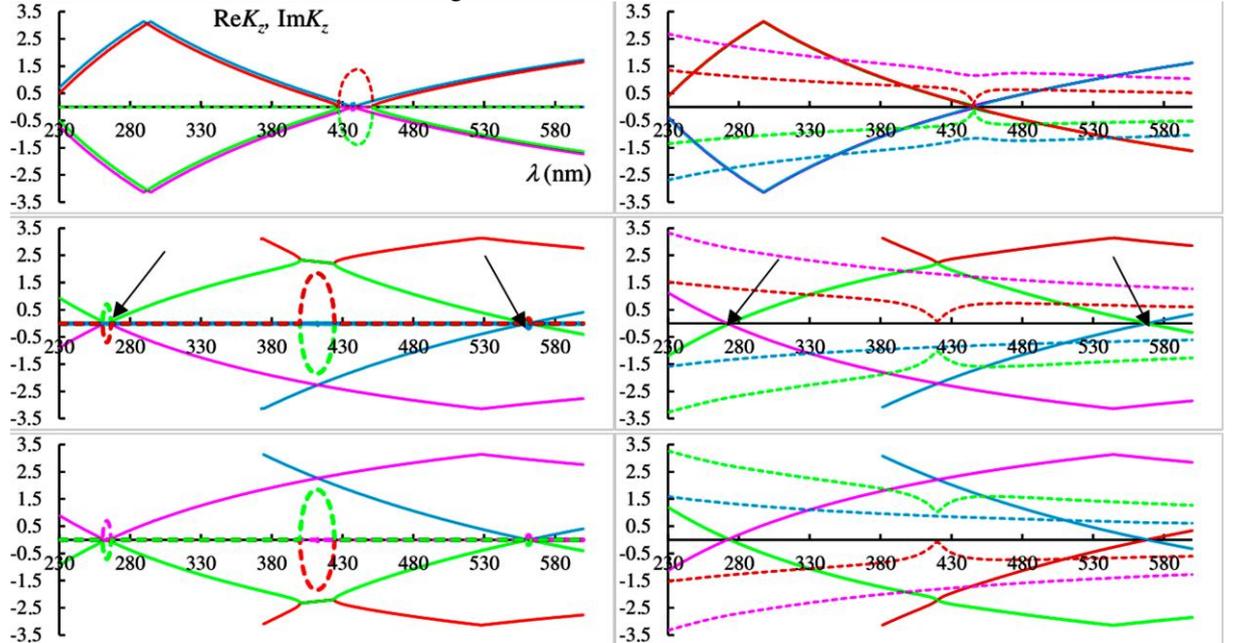

Fig. 11. Dependencies of real (solid lines) and imaginary (dashed lines) parts of Bloch wave vectors on wavelength in the absence of an external magnetic field (first row), in the presence of an external magnetic field at g=1 (second row), and at g=−1 (third row) at an angle $\vartheta = 45°$. The right column corresponds to dichroic CLC. The left column shows the same dependencies for conventional CLC with parameters $\text{Re}\varepsilon_1 = 2.29$, $\text{Re}\varepsilon_2 = 2.143$, $\text{Im}\varepsilon_1 = \text{Im}\varepsilon_2 =$

0. The red and green lines correspond to resonant wave vectors, and the blue and purple lines correspond to non-resonant wave numbers.

Fig. 12 shows the reflection spectra in the absence of an external magnetic field, i.e., at g=0 (first row), in the presence of an external magnetic field at g=1 (second row), and at g=−1 (third row) in the case of oblique light incidence at an angle $\theta = 45°$. The right column corresponds to dichroic CLC. For comparison, the left column shows the same spectra for conventional CLC with parameters $\text{Re}\varepsilon_1 = 2.29$, $\text{Re}\varepsilon_2 = 2.143$. The incident light has polarization coinciding with the first EP (blue line) and with the second EP (red line). The case of minimal influence of dielectric boundaries $n_s = \sqrt{\varepsilon_0}$ is considered.

As can be seen from the graphs, new PBGs are formed in the areas around the lateral Dirac points for conventional CLCs (they are indicated by arrows), while for dichroic CLCs there is a slight increase in reflection near the short-wave Dirac point and a dip in reflection near the long-wave Dirac point (they are also indicated by arrows). If at g=1 near the short-wave Dirac point, the diffraction reflection undergoes light with polarization coinciding with the first EP, as in the main PBG, then near the long-wave Dirac point, the diffraction reflection undergoes light with polarization coinciding with the second EP. At g= −1, we have the opposite picture.

Thus, when light is incident at an angle in the presence of an external magnetic field, conventional CLCs exhibit a complex band structure. In this case, not one but three PBGs arise (as shown by our numerical simulations, even at small angles of incidence, when the splitting of the main PBG [17] has not yet occurred. The central (main) PBG is determined by the structure of the CLC and exists both in the absence of an external magnetic field and in its presence, and moreover, both under normal light incidence and under oblique light incidence. The two new side PBGs are also sensitive to the polarization of the incident light, as is the main one. However, while for the main PBG the chirality sign of the polarization of the incident diffracting light is determined only by the chirality sign of the CLC helix, for the new side PBGs it is determined depending on whether the direction of the external magnetic field and the direction of the incident light form an acute or obtuse angle.

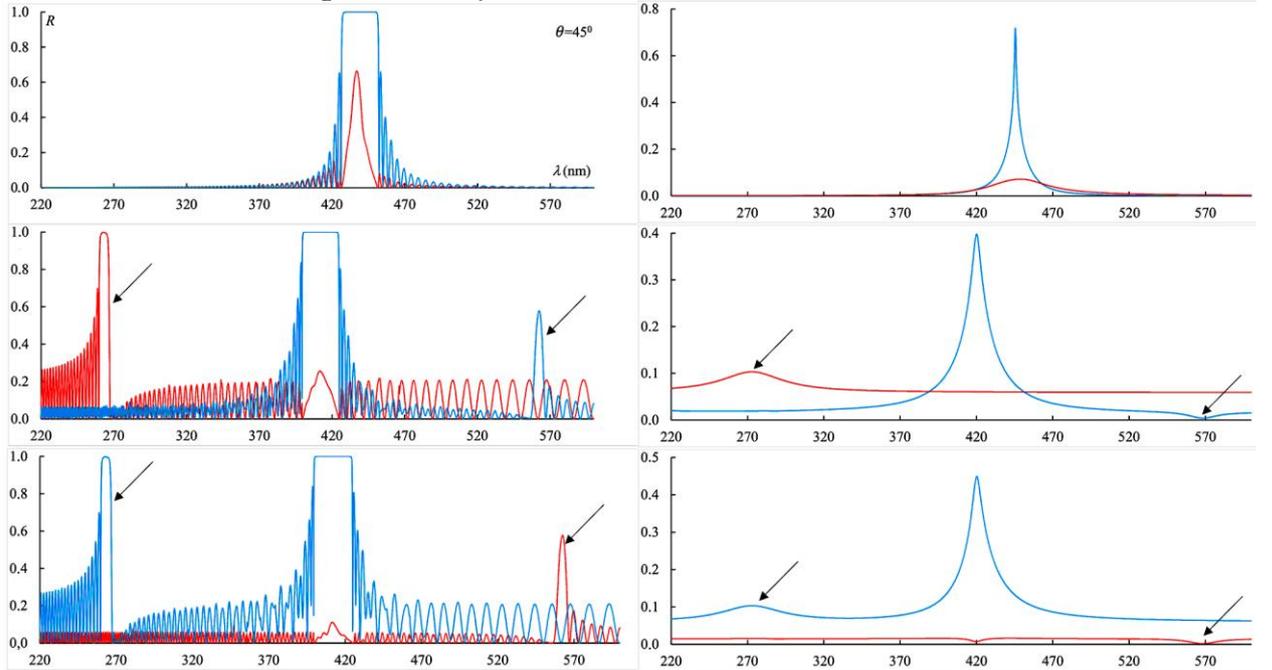

Fig. 12. Reflection spectra in the absence of an external magnetic field, i.e., at g=0 (first row), in the presence of an external magnetic field at g=1 (second row), and at g= −1 (third row) in the case of oblique light incidence at an angle of incidence $\theta = 45°$. The right column corresponds to dichroic CLC. For comparison, the left column shows the same spectra for conventional CLC with parameters $\text{Re}\varepsilon_1 = 2.29$, $\text{Re}\varepsilon_2 = 2.143$, $\text{Im}\varepsilon_1 = \text{Im}\varepsilon_2 = 0$. The incident light has polarization coinciding with the first EP (blue curve) and with the second EP (red curve). The case of minimal influence of dielectric boundaries $n_s = \sqrt{\varepsilon_0}$ is considered.

### 3.3. The effect of an external magnetic field on EP.

Let us now turn to the study of the properties of EPs of dichroic CLCs in an external magnetic field. As noted above, in the absence of an external magnetic field for conventional CLCs, EP polarization shifts from circular to linear polarization as the angle of incidence increases. Fig. 13 shows the EP ellipticity spectra for dichroic CLCs in the absence of an external magnetic field (at g=0; first column), in the presence of an external magnetic field at g=1 (second column), and at g= −1 (third column) and at different angles of incidence, namely at $\theta = 0°$ (first row), $\theta = 30°$ (second row), and $\theta = 60°$ (third row) in the case of minimal influence of dielectric boundaries $n_s = \sqrt{\varepsilon_0}$. As can be seen from the figure, when light is incident at an angle in the region near the Bragg frequency, there is a significant increase in the absolute values of the EPs ellipticities. With a further increase in the angle of incidence, there is a slight decrease in the absolute values of the EPs ellipticities. In the presence of an external magnetic field, there is practically no change in the ellipticities of EPs when the light is incident at an angle. Thus, the external magnetic field suppresses the shift of EPs from circular to linear polarizations as the angle of incidence increases.

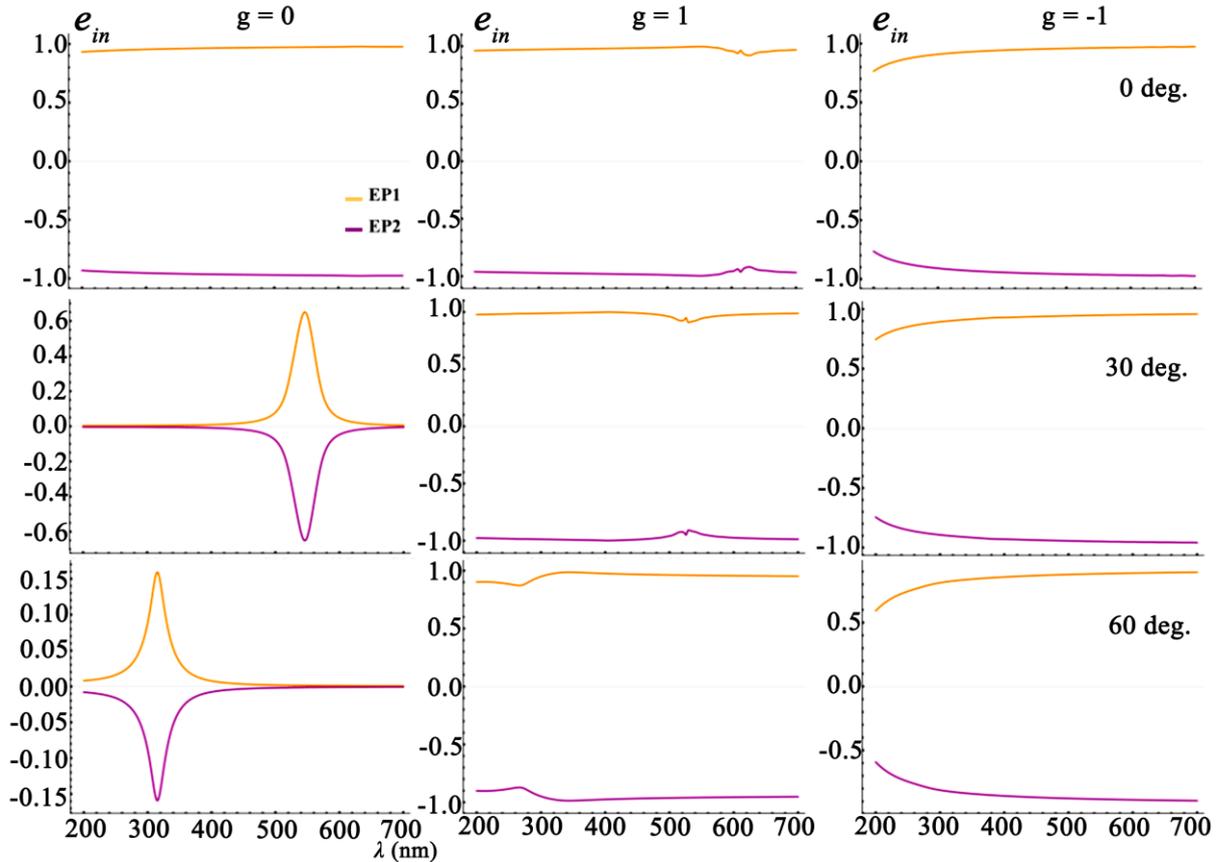

Fig. 13. Spectra of EPs ellipticities for dichroic CLC in the absence of an external magnetic field (at g=0; first column), in the presence of an external magnetic field at g=1 (second column) and at g= −1 (third column) and at different angles of incidence, namely at $\theta = 0°$ (first row), at $\theta = 30°$ (second row), and at $\theta = 60°$ (third row) in the case of minimal influence of dielectric boundaries $n_s = \sqrt{\varepsilon_0}$.

## 5. CONCLUSIONS

Thus, in this work, we investigated the magneto-optical properties of dichroic CLCs. We revealed unique properties of such structures, namely, we showed that at large angles of incidence in the case of $n_s = 1$, the resonance peak of the diffraction reflection is replaced by a resonance dip, and at the same time, a resonant increase in transmission occurs, i.e., resonant Bragg diffraction transmission appears instead of diffraction reflection (and this occurs at grazing angles of incidence). Thus, constructive interference of reflected secondary waves is replaced by destructive interference, and, conversely, destructive interference of transmitted secondary waves is replaced by constructive interference. In the case of $n_s = \sqrt{\varepsilon_0}$, higher orders of diffraction reflection appear again at oblique incidence of light at grazing angles. It is shown that at oblique incidence of light in the presence of an external magnetic field, new Dirac points appear in the spectra of such structures at each order of diffraction reflection. Furthermore, it is shown that an external magnetic field suppresses the shift of the polarization of eigenmodes from circular to linear polarization as the angle of incidence increases.

In conclusion, we note the importance of the results obtained here in connection with the design, synthesis, and characterization of a novel series of conjugated H-shaped liquid crystal molecules with exceptionally low birefringence [108].

## ACKNOWLEDGMENTS
The study was supported by the Ministry of Science and Higher Education of the Russian Federation within the projects FZNS-2023-0012 and № 075-03-2025-421.